\documentclass[conference]{IEEEtran}
\IEEEoverridecommandlockouts
\usepackage{cite}
\usepackage{amsmath,amssymb,amsfonts}
\usepackage{algorithmic}
\usepackage{graphicx}
\usepackage{textcomp}
\usepackage{xcolor}
\def\BibTeX{{\rm B\kern-.05em{\sc i\kern-.025em b}\kern-.08em
    T\kern-.1667em\lower.7ex\hbox{E}\kern-.125emX}}

\usepackage{subcaption}
\usepackage{tikz}
\usetikzlibrary{quantikz2}
\usepackage{comment}
\usepackage{booktabs}

\begin{document}

\title{An Evaluation of the Remote CX Protocol under Noise in Distributed Quantum Computing
\thanks{© 2026 IEEE.  Personal use of this material is permitted.  Permission from IEEE must be obtained for all other uses, in any current or future media, including reprinting/republishing this material for advertising or promotional purposes, creating new collective works, for resale or redistribution to servers or lists, or reuse of any copyrighted component of this work in other works.

Sponsored in part by the Bavarian Ministry of Economic Affairs, Regional Development and Energy as part of the 6GQT project.}
}

\author{\IEEEauthorblockN{Leo Sünkel}
\IEEEauthorblockA{\textit{Institute for Informatics} \\
\textit{LMU Munich}\\
Munich, Germany \\
leo.suenkel@ifi.lmu.de}
\and
\IEEEauthorblockN{Michael Kölle}
\IEEEauthorblockA{\textit{Institute for Informatics} \\
\textit{LMU Munich}\\
Munich, Germany}
\and
\IEEEauthorblockN{Tobias Rohe}
\IEEEauthorblockA{\textit{Institute for Informatics} \\
\textit{LMU Munich}\\
Munich, Germany}
\and
\IEEEauthorblockN{Claudia Linnhoff-Popien}
\IEEEauthorblockA{\textit{Institute for Informatics} \\
\textit{LMU Munich}\\
Munich, Germany}
}

\maketitle

\begin{abstract}
Quantum computers connected through classical and quantum communication channels can be combined to function as a single unit to run large quantum circuits that each device is unable to execute on their own. The distributed quantum computing paradigm is therefore often seen as a potential pathway to scaling quantum computing to capacities necessary for practical and large-scale applications. Whether connecting multiple quantum processing units (QPUs) in clusters or over networks, quantum communication requires entanglement to be generated and distributed over distances. 
Using entanglement, the remote CX protocol can be performed, which allows the application of the CX gate involving qubits located in different QPUs. In this work, we use a specialized simulation framework for a high-level evaluation of the impact of the protocol when executed under noise in various network configurations using different number of QPUs. We compare naive and graph partitioning qubit assignment strategies and how they affect the fidelity in experiments run on Grover, GHZ, VQC, and random circuits. The results provide insights on how QPU and network configurations or naive scheduling can degrade performance.
\end{abstract}


\section{Introduction}
Distributed quantum computing \cite{buhrman2003distributed,van2016path,caleffi2024distributed} on quantum communication networks is an emerging paradigm in which circuits are executed on multiple connected quantum processing units (QPUs). However, enabling distributed quantum computing on large networks poses significant challenges \cite{cacciapuoti2019quantum,kozlowski2019towards,caleffi2018quantum}, especially for networks akin to a quantum Internet \cite{kimble2008quantum,azuma2023quantum}. Entanglement is a central component of such networks and is required for quantum communication. Unfortunately, qubits are fragile, and establishing and maintaining entanglement over distance for a prolonged time is a major challenge that requires specialized hardware with the ability to run certain communication protocols. To this end, quantum repeaters \cite{briegel1998quantum,dur1999quantum,van2013designing} performing entanglement swapping and purification protocols can be placed throughout the network, thereby enabling quantum communication over larger distances. Entanglement swapping allows the establishment of entanglement between nodes without requiring them to interact directly with each other, while the purification protocol can increase entanglement fidelity.
Providing the hardware to enable distributing and maintaining highly entangled qubits is not the only challenge; novel algorithms and software tools are also being heavily investigated by the research community. For instance, qubits must be assigned to nodes in the network and circuits compiled accordingly \cite{ferrari2021compiler,cuomo2023optimized,mao2023qubit,burt2024generalised}. Furthermore, algorithms and applications are also being evaluated, including chemistry \cite{jones2024distributed}, variational quantum algorithms \cite{khait2023variational,sunkel2025evaluating}, and a distributed version of Shor's algorithm \cite{yimsiriwattana2004distributed}. 
Performing remote operations between QPUs requires entanglement, which must be established and distributed in the network, and is ultimately consumed in the process. Thus, entanglement is often considered a key resource that should not be used wastefully. Naive qubit assignment and circuit partitioning may result in unnecessary remote operations, exhausting limited resources which could otherwise be avoided. The network configuration, i.e., what type of QPUs and how they are connected, or the circuit architecture may also be a contributing factor and influence the performance. 

The aim of this work is to evaluate how distributed versions of circuits perform under noise in comparison to their original monolithic counterparts. The distributed versions of the circuits are constructed for different network architectures, that is, different number of QPUs as well as qubit capacities. However, the simulations use a simplified model; more specifically, we do not use a full network simulator that applies networking protocols such as entanglement swapping or purification, nor does the model contain repeaters or entanglement generation over distance. Instead, the focus lies on a high-level evaluation of how remote operations, i.e., remote CX gates, which are added as a protocol to the circuit, influence the overall performance of the circuit measured in fidelity. 
In quantum networking, QPUs can reserve qubits for computation and communication, where the former can be seen as the logical qubits in a monolithic circuit and the latter are only used for communicating between QPUs. Different configurations (e.g., more communication qubits allow for more remote operations to be applied in parallel) result in a different distributed circuit, and thus required remote operations and resources. To this end, we investigate how different QPU configurations have an impact on the overall computation. In an experimental evaluation, we construct distributed circuits according to their QPU configuration and simulate their execution under noise. We additionally evaluate how the qubit schedule, that is, the assignment of qubits to QPUs affects the performance.

\section{Background}\label{sec:background}
We begin this section with a very short overview of the essentials of quantum computing and continue with quantum networks in the same way. Finally, we recapitulate the basics of distributed quantum computing.

\subsection{Quantum Computing}
In quantum computing (QC), the most fundamental unit of information is encoded in a qubit, a two-state system similar to a bit in classical computing. However, unlike a classical bit, a qubit can be in a superposition (i.e., a linear combination) of states, whereas a classical bit is either in state 0 or 1 at any given time \cite{nielsen2010quantum}. The state of a qubit can be defined as $\ket{\psi} = \alpha\ket{0} + \beta\ket{1}$ with $\alpha$ and $\beta$ being complex numbers representing probability amplitudes where $|\alpha|^2$ and $|\beta|^2$ give the probabilities of the qubit being in state $0$ or $1$ after measurement where $|\alpha|^2 + |\beta|^2=1$ \cite{nielsen2010quantum}. That is, a qubit is in a quantum state until it's measured; after measurement the qubit collapses to the corresponding basis state. A further important property is entanglement. When qubits are entangled, their measurement outcomes correlate, that is, by measuring one qubit, one can infer information about the others, even if the qubits are far apart. Bell states are maximally entangled two qubit states and are defined as $\Phi^+ = \frac{1}{\sqrt{2}}(\ket{00} + \ket{11})$,$\Phi^- = \frac{1}{\sqrt{2}}(\ket{00} - \ket{11})$, $\Psi^+ = \frac{1}{\sqrt{2}}(\ket{01} + \ket{10})$, and $\Psi^- = \frac{1}{\sqrt{2}}(\ket{01} - \ket{10})$ \cite{nielsen2010quantum}. The qubits of these states are often referred to as Bell or EPR pairs, and are ubiquitous in quantum networking.

\subsection{Quantum Networks}
\begin{figure}[tb]
    \centering
    \scalebox{0.8}{
        \begin{quantikz}
            \lstick{\ket{\psi_0}} & \ctrl{1} & \gate{H} & \meter{}\wire[d][2]{c} && \\
            \lstick[2]{$\Phi^+$} & \targ{} &&& \meter{}\wire[d][1]{c} & \\
            &&& \gate{X} & \gate{Z} & \rstick[1]{$\ket{\psi_1}$}
        \end{quantikz}
    }
    \caption{The teleportation protocol \cite{caleffi2024distributed}. The state $\psi_0$ is transferred to qubit $\psi_1$, which results in its destruction at its original location. The protocol requires a third party and a Bell pair.}
    \label{fig:teleportation_protocol}
\end{figure}
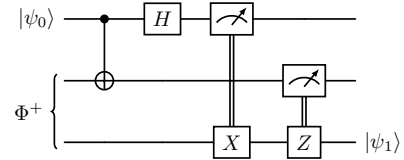

A quantum network can be defined as a graph $G=(N, E)$ with nodes $N$ and edges $E$ where $N$ are quantum devices (e.g., QPUs or repeaters) and $E$ communication channels. Note that both quantum and classical channels are required for certain protocols. The crucial requirement for quantum communication is entanglement, which must be created and distributed throughout the network. However, to establish and maintain entanglement over large distances, quantum repeaters performing entanglement purification and swapping are required. By applying a purification protocol on $n$ weakly entangled qubits, $m$ highly entangled qubits can be established, where $m<n$. With the entanglement swapping protocol, entanglement between two nodes, which have not necessarily directly interacted with each other, can be established. The protocol involves two Bell pairs and three parties. Fig. \ref{fig:entanglement_swapping} shows a high-level overview. Nodes A and B each share a Bell Pair with C (depicted left in the figure). A Bell measurement can now be performed on the qubits located at C, thereby projecting the entanglement onto A and B, and thus establishing an entangled link between them without requiring direct communication between the nodes.
A qubit's state can be teleported to different nodes in the network. More specifically, the quantum teleportation protocol requires a Bell pair established between two QPUs and allows one to transfer a qubit's state to another. The state, however, is destroyed at its original location. The protocol is shown in Fig. \ref{fig:teleportation_protocol}.

\begin{figure}[tb]
    \centering
    \scalebox{0.8}
    {
        \begin{tikzpicture}
            \node[draw, circle, fill=blue!20](A) at (0, 0) {A};
            \node[draw, circle, fill=green!20](B) at (2, 0) {B};
            \node[draw, circle, fill=red!20](C) at (1, 1) {C};
            \draw[-, very thick] (A.north) -- node[midway, above] {$\Phi^+$} (C.west);
            \draw[-,blue, very thick] ([xshift=4pt,yshift=-1pt]A.north) -- ([xshift=1.4pt,yshift=-4pt]C.west);
            
            \draw[-, very thick] (B.north) -- node[midway, above, xshift=4pt] {$\Phi^+$} (C.east);
            \draw[-,blue, very thick] ([xshift=-4pt,yshift=-1pt]B.north) -- ([xshift=-1.35pt,yshift=-4pt]C.east);

            \draw[-{Latex[width=2.5mm]}, ultra thick] (2.5, 0.5) -- node[midway, above]{ES protocol} (4.5, 0.5);
    
            \node[draw, circle, fill=blue!20](A) at (5, 0) {A};
            \node[draw, circle, fill=green!20](B) at (7, 0) {B};
            \node[draw, circle, fill=red!20](C) at (6, 1) {C};
            
            \draw[-, very thick] (A.east) -- node[midway, above] {$\Phi^+$} (B.west);
            
        \end{tikzpicture}
    }
    \caption{The entanglement swapping (ES) protocol requires three parties and two Bell pairs, as well as classical communication (blue lines). A and C, and B and C each share a Bell pair. C applies a Bell state measurement and sends the results over a classical communication channel to A and B who each can then apply correcting gates.  By applying this protocol, the entanglement can be transferred such that A and B become entangled.}
    \label{fig:entanglement_swapping}
\end{figure}
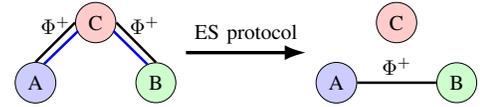

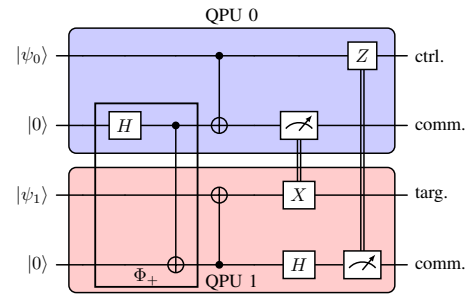
\begin{figure}[tb]
 \centering
 \scalebox{0.7}{
     \begin{quantikz}[row sep=0.8cm]
         \lstick{\ket{\psi_0}} & \gategroup[2, steps=7, style={rounded corners, fill=blue!20}, background, label style={label position=above}]{QPU 0} &&& \ctrl{1} &&& \gate{Z} & \rstick[1,brackets=none]{ctrl.} \\
         \lstick{\ket{0}} && \gate{H} \gategroup[3, steps=2, style={inner ysep=0.9pt, line width=1pt}, label style={label position=below}]{$\Phi_+$} & \ctrl{2} & \targ{} && \meter{}\wire[d][1]{c} && \rstick[1,brackets=none]{comm.}  \\
         \lstick{\ket{\psi_1}} & \gategroup[2, steps=7, style={rounded corners, fill=red!20}, background, label style={label position=below}]{QPU 1} &&& \targ{} && \gate{X} && \rstick[1,brackets=none]{targ.}  \\
         \lstick{\ket{0}} &&& \targ{} & \ctrl{-1} && \gate{H} & \meter{}\wire[u][3]{c} & \rstick[1,brackets=none]{comm.} 
     \end{quantikz}
 }
 \caption{The remote CX protocol \cite{ferrari2021compiler}. The protocol requires a Bell pair between a communication qubit of each QPU. }
 \label{fig:remote_cx}
\end{figure}

\begin{figure*}[tb]
    \centering
    \begin{subfigure}[b]{0.45\textwidth}
        \centering
        \scalebox{0.75}
        {
        \raisebox{1\height}
        {
            \begin{quantikz}
                \lstick{\ket{\psi_0}} & \gategroup[2, steps=6, style={rounded corners, fill=blue!20}, background]{} & \gate{X} & \ctrl{1} &&& \meter{} & \rstick[2]{QPU 0} \\
                \lstick{\ket{\psi_1}} &&& \targ{} & \ctrl{1} \gategroup[2, steps=1, style={rounded corners}]{RCX}  && \meter{} &\\
                \lstick{\ket{\psi_2}} & \gategroup[2, steps=6, style={rounded corners, fill=red!20}, background]{} &&& \targ{} & \ctrl{1} & \meter{} & \rstick[2]{QPU 1} \\
                \lstick{\ket{\psi_3}} &&&&& \targ{} & \meter{} &
            \end{quantikz}
        }
        }
        \caption{An example circuit where qubits marked blue are assigned to QPU 0 and red qubits to QPU 1. The circuit contains an remote operation (remote CX) indicated by the RCX box.}
        \label{fig:example_qc_2}
    \end{subfigure}
    \hfill
    \begin{subfigure}[b]{0.45\textwidth}
        \centering
        \scalebox{0.7}
        {
        \begin{quantikz}[row sep=0.8cm]
            \lstick{\ket{\psi_0}} \gategroup[2, steps=10, style={rounded corners, fill=blue!20, inner xsep=-1.pt}, background, label style={label position=mid}]{} & \gate{X} & \ctrl{1} &&&&&&& &\rstick[4]{Comp.} \\
            \lstick{\ket{\psi_1}} && \targ{} & \gategroup[5, steps=6,style={inner ysep=-1pt, line width=2pt}]{Remote CX} & && \ctrl{3} & \gate{Z} &&& \\
            \lstick{\ket{\psi_2}} \gategroup[2, steps=10, style={rounded corners, fill=red!20, inner xsep=-1.pt}, background, label style={label position=mid}]{} &&&&& \targ{} &&& \gate{X} & \ctrl{1} &  \\
            \lstick{\ket{\psi_3}} &&&&&&&&& \targ{} & \\
            \lstick{\ket{0}} \gategroup[1, steps=10, style={rounded corners, fill=blue!20, inner xsep=-1.pt}, background, label style={label position=mid}]{} &&& \gate{H} & \ctrl{1} & & \targ{} && \meter{}\wire[u][2]{c} && \rstick[2]{Comm.} \\
            \lstick{\ket{0}} \gategroup[1, steps=10, style={rounded corners, fill=red!20, inner xsep=-1.pt}, background, label style={label position=mid}]{} &&&& \targ{} & \ctrl{-3} & \gate{H} & \meter{}\wire[u][4]{c} &&& 
        \end{quantikz}
        }
        \caption{The distributed version of the left circuit. The remote CX protocol is inserted in place of the RCX marked in the monolithic circuit.}
        \label{fig:example_dqc_circuit}
    \end{subfigure}
    \caption{On the left, the circuit only with computational qubits, i.e, the architecture of the monolithic circuit is shown. Qubits marked blue are assigned to QPU 0 and red indicates QPU 1. On the right, the distributed version of the circuit is shown, including computational and communication qubits.}
    
\end{figure*}
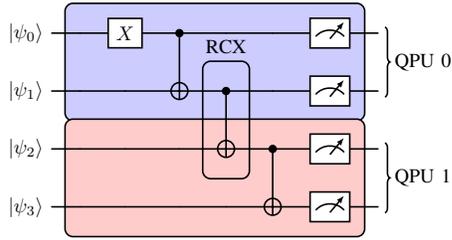
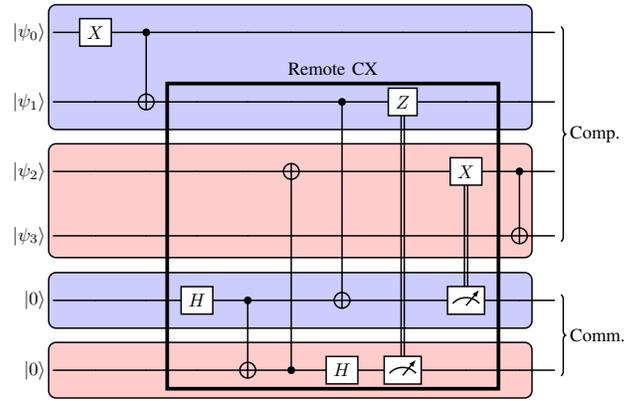

\subsection{Distributed Quantum Computing}
In distributed quantum computing (DQC), a quantum circuit is run on multiple connected QPUs simultaneously. Qubits are assigned to QPUs and the circuit is divided into subcircuits which are then distributed to the nodes (i.e., QPUs). When, for example, a CX gate where the involved qubits are located in different QPUs must be executed, a remote operation must be performed. One possible solution is to transfer the state of a qubit through the teleportation protocol (shown in Fig. \ref{fig:teleportation_protocol}) such that all necessary qubits are located in the same QPU; the gate can then be executed locally. Another option is to use the remote CX protocol, which is shown in Fig. \ref{fig:remote_cx}. In this approach, the qubits do not change their location. Both protocols require entanglement in the form of a Bell pair. However, which protocol to use depends on the context.

\section{Related Work}
In \cite{parekh2021quantum} the authors present a simulation framework and evaluate different quantum algorithms in the DQC setting, including the variational quantum eigensolver and quantum k-means clustering. Another DQC simulation framework is presented in \cite{muralidharan2025simulation}; here the authors also evaluate their approach with various algorithms. A performance analysis of DQC is given in \cite{mao2024performance}, variational quantum eigensolvers for DQC are discussed in \cite{khait2023variational} and \cite{diadamo2021distributed}. The architecture of variational quantum circuits in the context of DQC was evaluated in \cite{sunkel2025evaluating}; the simulation framework used in this work is based on the approach presented in that paper. DQC for chemical applications are discussed in \cite{jones2024distributed,jones2025analyzing}.

\section{Approach}\label{sec:approach}
In the first part of this section, we introduce the approach used to construct and simulate the distributed circuits. In the second part, we give an overview of the experimental setup.

\subsection{Simulating a Distributed Circuit}
The number of available QPUs and their respective capabilities (i.e., number of computational qubits) influence the required number of remote operations. This in turn can affect the performance as more non-local operations demand creating, distributing, and maintaining highly entangled qubits. Moreover, increasing the number of operations also results in more noise and decoherence. The aim of this work is to investigate how the number of QPUs and their capabilities affect the circuit's performance under noise in a DQC setting. To this end, we evaluate a number of different circuits under various conditions and our simulation approach is as follows. Given a monolithic circuit and a network configuration that specifies the number of QPUs and their respective number of computational and communication qubits, as well as a qubit assignment to QPUs, we create a larger circuit that incorporates the qubits of all QPUs. For example, a monolithic circuit with four qubits and a network configuration with two QPUs with each having two computational and one communication qubit would result in a circuit containing six qubits in total. Remote operations (i.e, CX gates) between QPUs are added as the remote CX protocol. Note that this is a simplified model and does not consider routers or entanglement swapping and purification. The aim is to evaluate how the performance is affected by the protocol alone under noise. Fig. \ref{fig:example_qc_2} depicts a circuit with one remote CX gate; the corresponding distributed version of the same circuit is illustrated in Fig. \ref{fig:example_dqc_circuit} where blue indicates qubits assigned to QPU 0 and red to QPU 1. The approach is based on the one introduced in \cite{sunkel2025evaluating}.  
In the experiments, circuits are executed under noise and the fidelity between the resulting and ideal state is calculated. The fidelity between quantum states can be used as a measure to determine how similar the states are; it is defined as $F(\rho, \sigma) = \textnormal{tr}\left(\sqrt{\sqrt{\rho}\sigma\sqrt{\rho}}\right)$ where $\rho$ and $\sigma$ are quantum states \cite{nielsen2010quantum}. The noise model and further configuration of the experiments are described next.

\begin{table}[tb]
    \centering
    \caption{Overview of the configurations used in this work. Each circuit was run for five different seeds. The computational and communication qubits listed in the table are for each QPU; the total qubits refers to the number of qubits in the entire simulated distributed circuit.}
    \begin{tabular}{ccccccc}
        \toprule
        ID & Circuit & Qubits & QPU & Comp. & Comm. & Total Qubits \\ 
        && (Logical) && Qubits & Qubits & \\ \midrule
        1 & Grover & 4 & 2 & 2 & 2 & 8\\ 
        2 & GHZ & 8 & 2 & 4 & 2 & 12 \\
        3 & GHZ & 8 & 4 & 2 & 2 & 16 \\
        4 & GHZ & 8 & 8 & 1 & 2 & 24 \\
        5 & Random & 12 & 4 & 3 & 1 & 16 \\
        6 & Random & 12 & 4 & 3 & 2 & 20 \\
        7 & Random & 12 & 4 & 3 & 3 & 24\\
        8 & Random & 8 & 2 & 4 & 2 & 12 \\
        9 & Random & 8 & 4 & 2 & 2 & 16 \\
        10 & Random & 8 & 8 & 1 & 2 & 24\\
        11 & Random & 12 & 2 & 6 & 2 & 16\\
        12 & Random & 12 & 3 & 4 & 2 & 18 \\
        13 & Random & 12 & 4 & 3 & 2 & 20 \\
        14 & VQC & 8 & 2 & 4 & 2 & 12 \\
        15 & VQC & 8 & 4 & 2 & 2 & 16 \\
        16 & VQC & 8 & 8 & 1 & 2 & 24 \\
        17 & VQC & 12 & 2 & 6 & 2 & 16 \\
        18 & VQC & 12 & 3 & 4 & 2 & 18 \\
        19 & VQC & 12 & 4 & 3 & 2 & 20 \\
         \bottomrule
    \end{tabular}
    \label{tab:circuit_overview}
\end{table}

\begin{figure*}[tb]
    \centering
    \begin{subfigure}[t]{0.3\textwidth}
        \centering
            \includegraphics[scale=0.28]{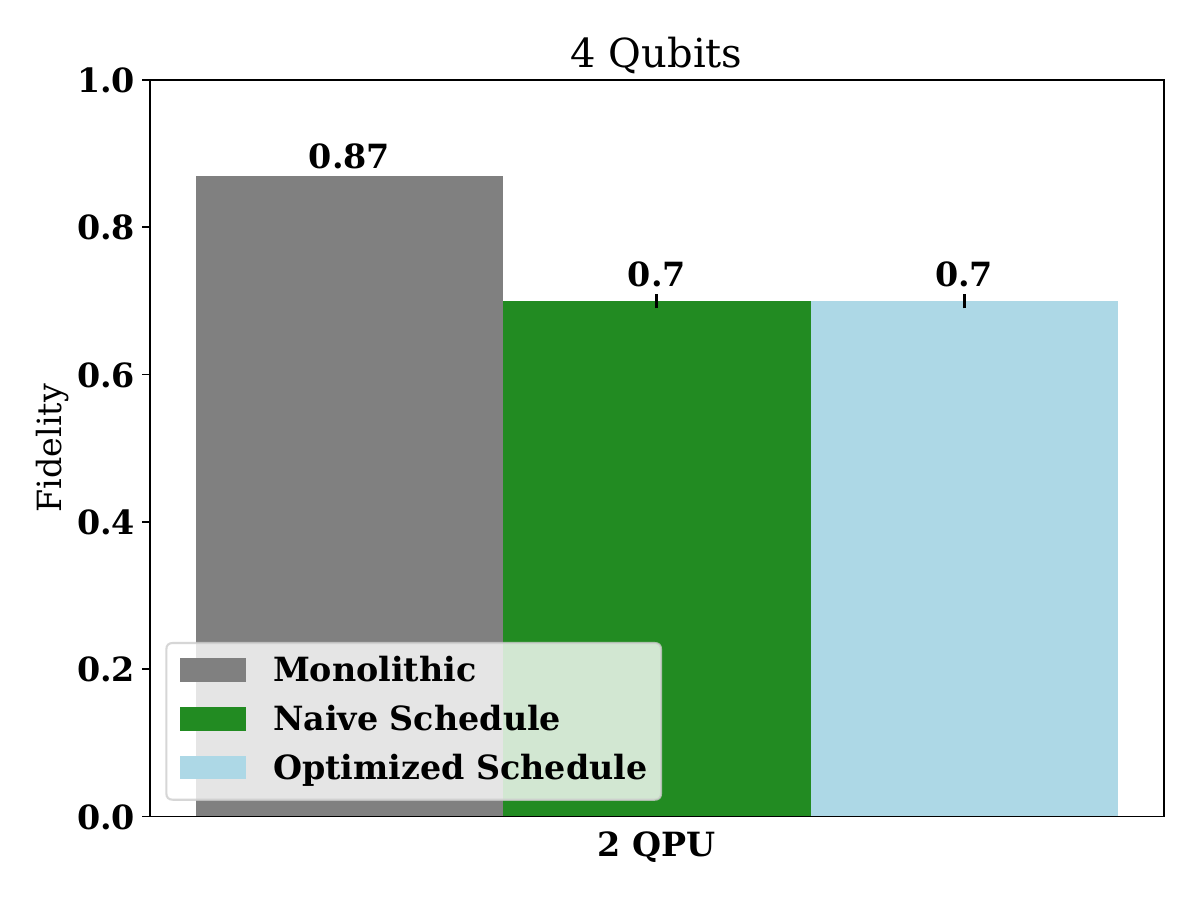}
        \caption{Grover circuit with four qibits.}
        \label{fig:grover_result}
    \end{subfigure}
    \hfill
    \begin{subfigure}[t]{0.3\textwidth}
        \centering
            \includegraphics[scale=0.28]{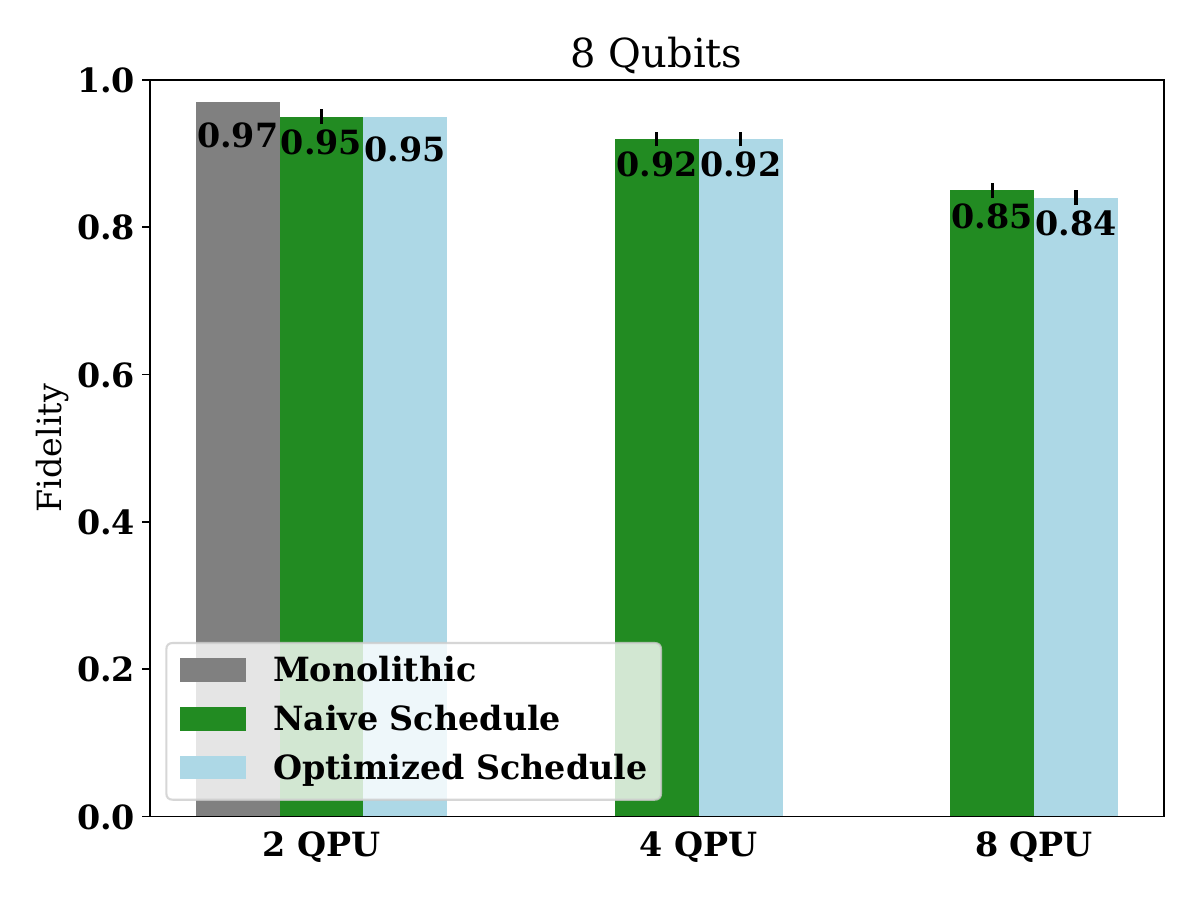}
        \caption{GHZ circuit with eight qubits.}
        \label{fig:ghz_result}
    \end{subfigure}
    \hfill
    \begin{subfigure}[t]{0.3\linewidth}
        \centering
            \includegraphics[scale=0.28]{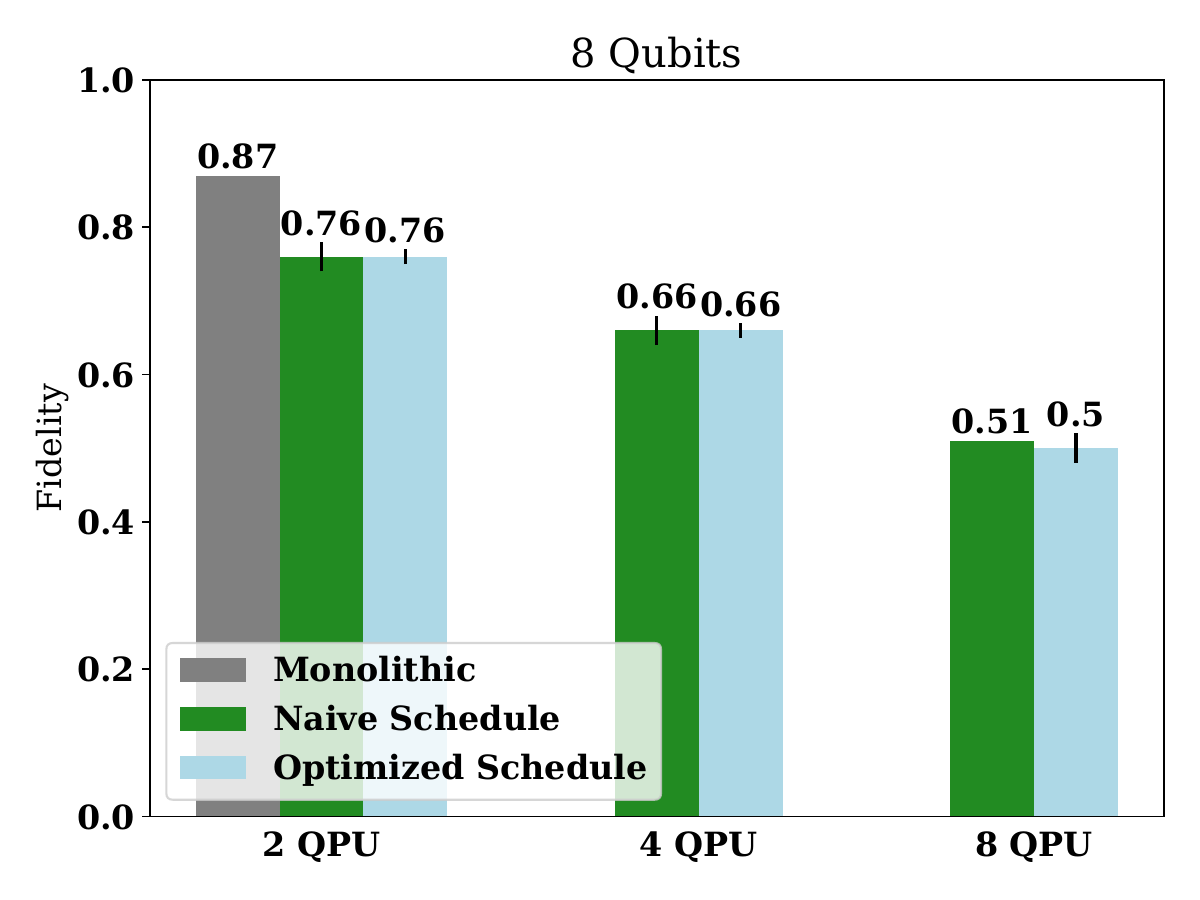}
        \caption{VQC with eight qubits.}
        \label{fig:vqc_8_results}
    \end{subfigure}
    \begin{subfigure}[t]{0.3\linewidth}
        \centering
            \includegraphics[scale=0.28]{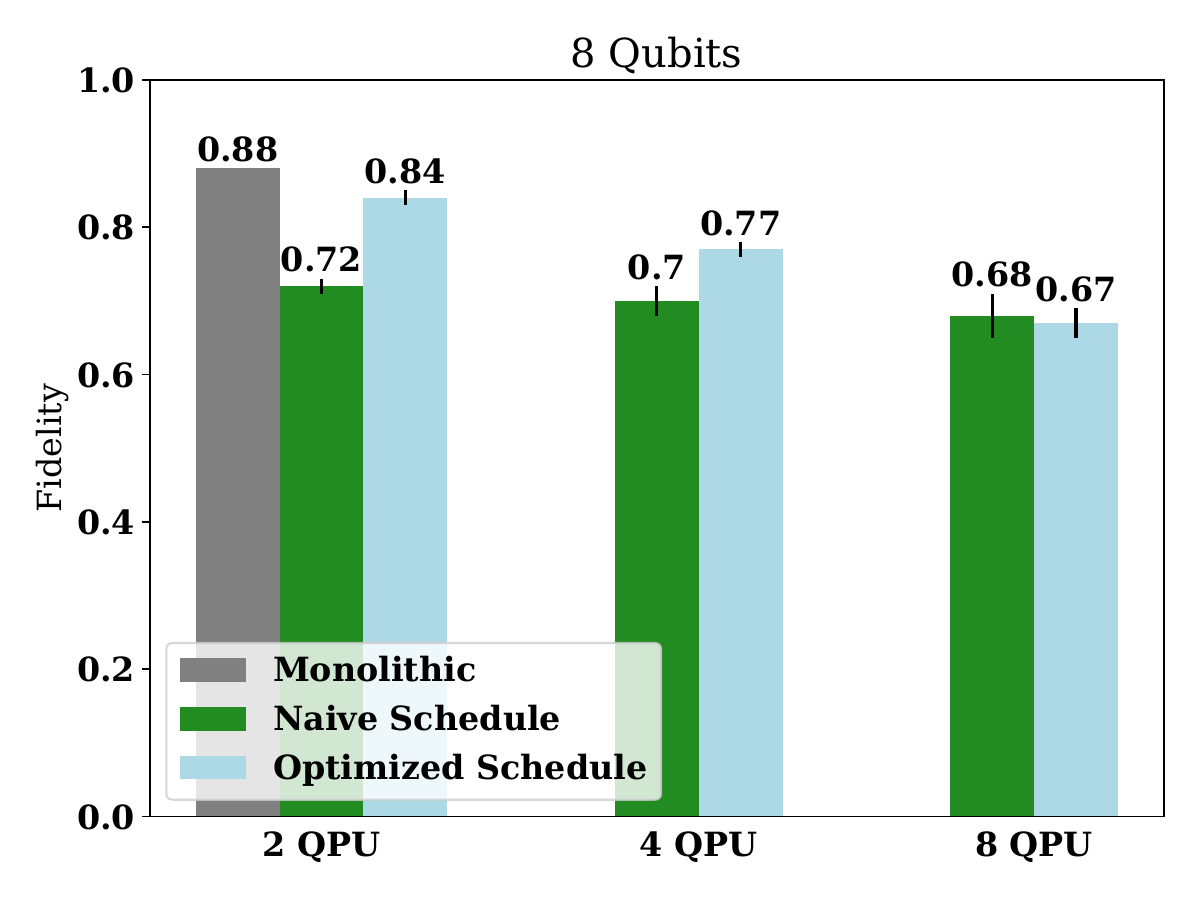}
        \caption{Random circuit with eight qubits.}
        \label{fig:random_8_results}
    \end{subfigure}
    \hfill
    \begin{subfigure}[t]{0.3\linewidth}
        \centering
            \includegraphics[scale=0.28]{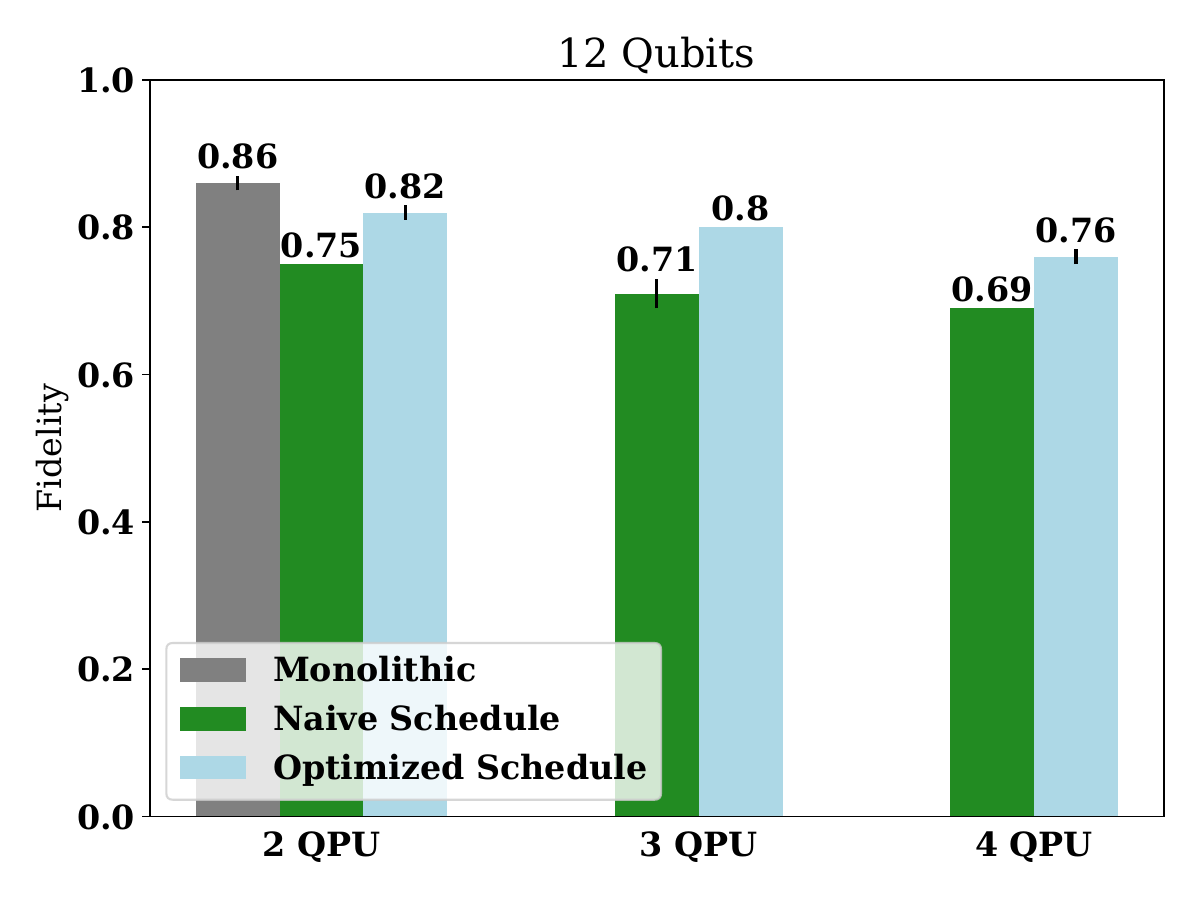}
        \caption{Random circuit with twelve qubits.}
        \label{fig:random_12_results}
    \end{subfigure}
    \hfill
    \begin{subfigure}[t]{0.3\textwidth}
        \centering
            \includegraphics[scale=0.28]{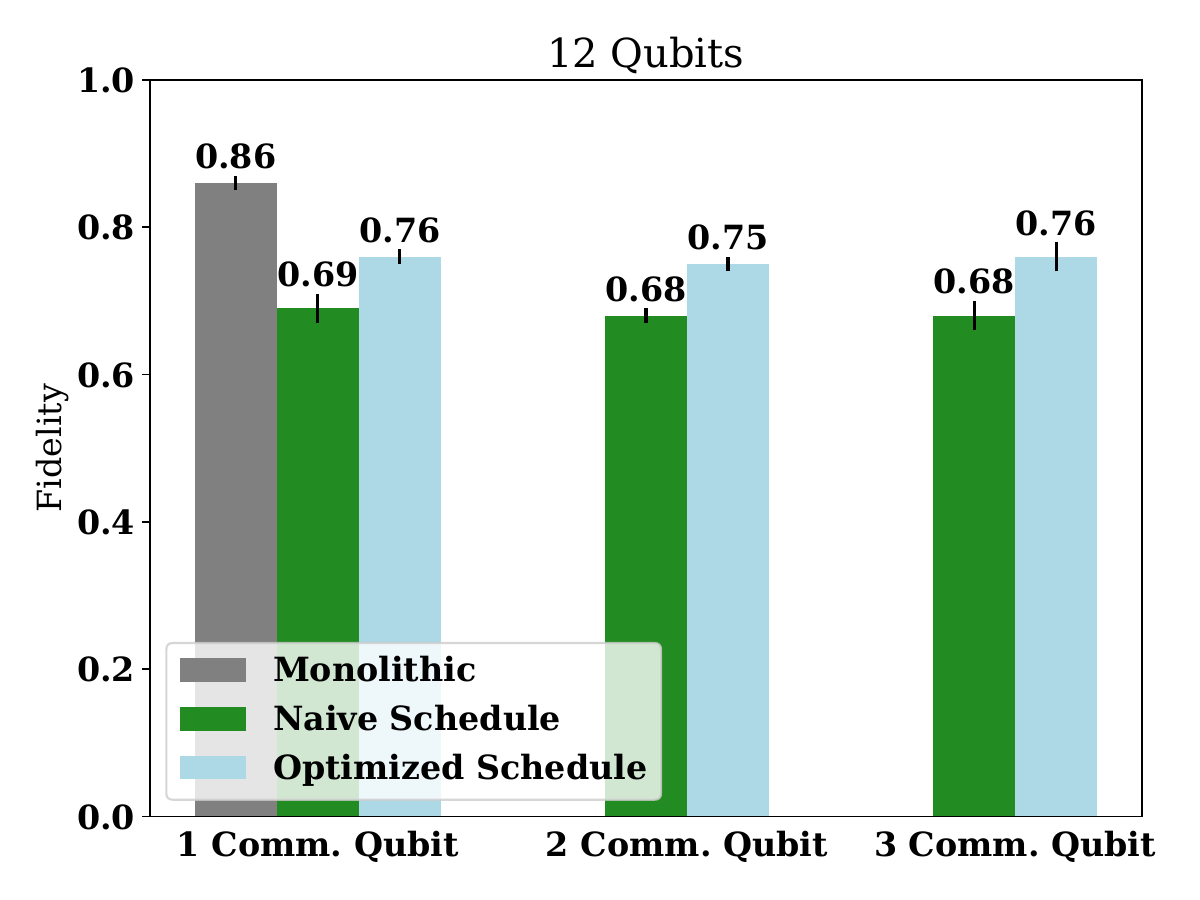}
        \caption{Random circuit on QPUs with increasing number of communication qubits.}
        \label{fig:comm_qubit_results}
    \end{subfigure}

    \begin{subfigure}[t]{0.3\textwidth}
        \centering
            \includegraphics[scale=0.28]{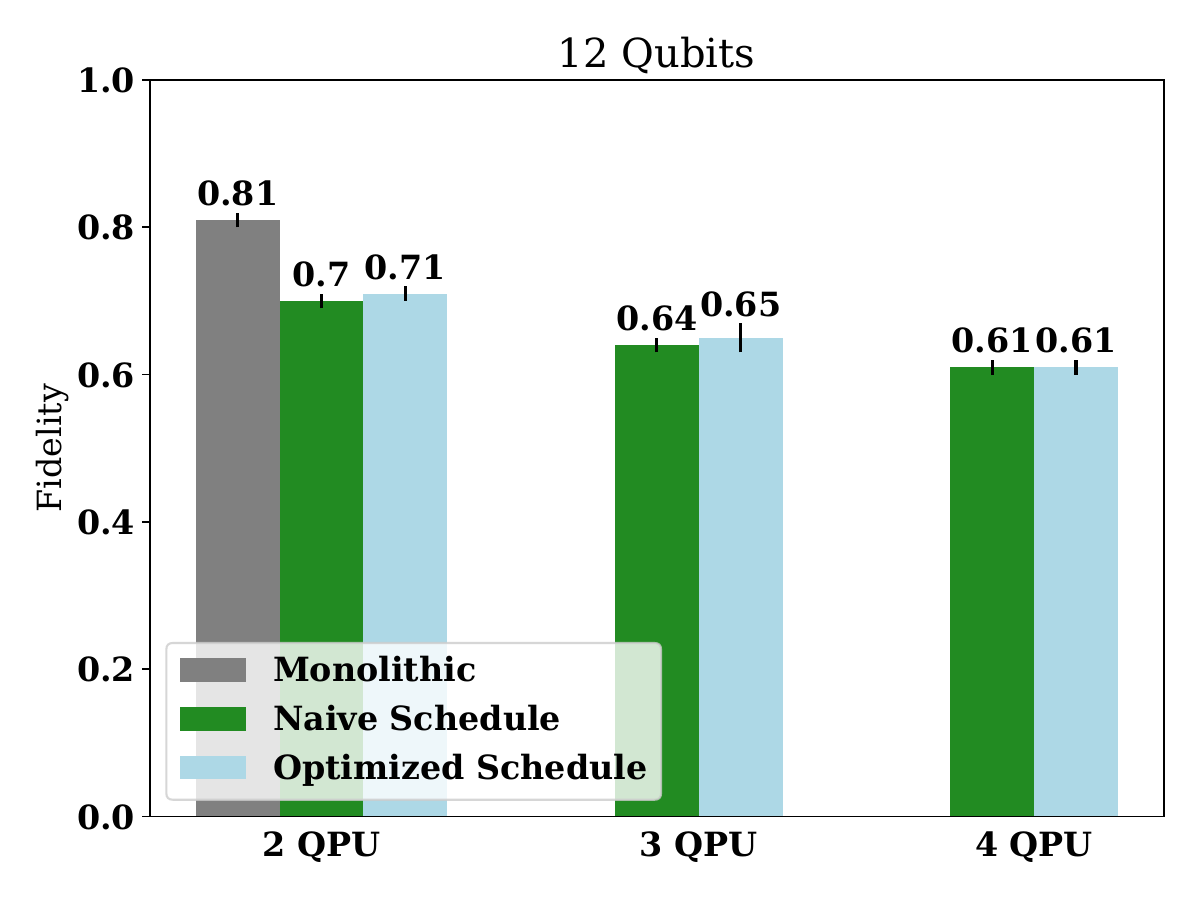}
        \caption{VQC with twelve qubits.}
        \label{fig:vqc_12_results}
    \end{subfigure}
    \caption{The results of all experiments. Runs are compared in terms of fidelity. All experiments were run for five different seeds. Plots show mean fidelity with std.}
    \label{fig:placeholder}
\end{figure*}

\subsection{Experimental Setup}
The approach is evaluated on Grover, GHZ, variational quantum circuits (VQCs), and random circuits on various QPU configurations. The custom noise mode uses depolarizing error on one and two-qubit gates; it also applies a readout error. The error probabilities are as follows: 0.001 for one qubit gates and 0.005 for two qubit gates and readout. All QPUs contain two communication qubits unless stated otherwise. Each experiment was run for five different seeds and we used Qiskit \cite{javadi2024quantum} to implement the circuits and run the simulations. An overview of the configurations used in the experiments is given in Table \ref{tab:circuit_overview}. All circuits were transpiled using the following basis gates: X, RZ, H, and CX. Furthermore, we evaluate the effect on the fidelity of two different qubit assignment methods. In the first technique, qubits were assigned to QPUs using a naive approach in which each QPU is successively filled up, that is, the first two qubits are assigned to QPU 0, the next two to QPU 1, and so forth. The second approach uses graph partitioning (GP) to assign qubits to QPUs (we refer to the resulting schedule as the optimized schedule below). In this technique, circuits are converted to graphs where the nodes are qubits and edges CX gates between the respective qubits weighted by the number of CX gates between them. Single-qubit gates are not represented in the graph. We use PyMetis for GP in this work.

\begin{figure*}[tb]
    \centering
    \begin{subfigure}[t]{0.3\textwidth}
        \centering
            \includegraphics[scale=0.35]{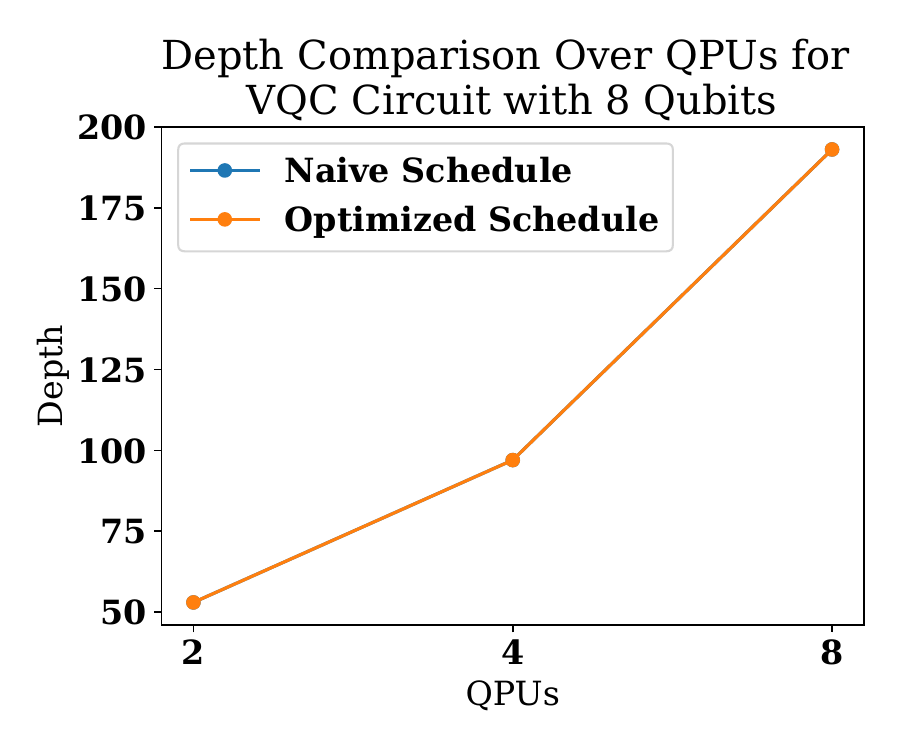}
        \caption{VQC depth comparison. Both scheduling methods result in identical circuit depth.}
        \label{fig:vqc_depth}
    \end{subfigure}
    \hfill
    \begin{subfigure}[t]{0.3\textwidth}
        \centering
            \includegraphics[scale=0.35]{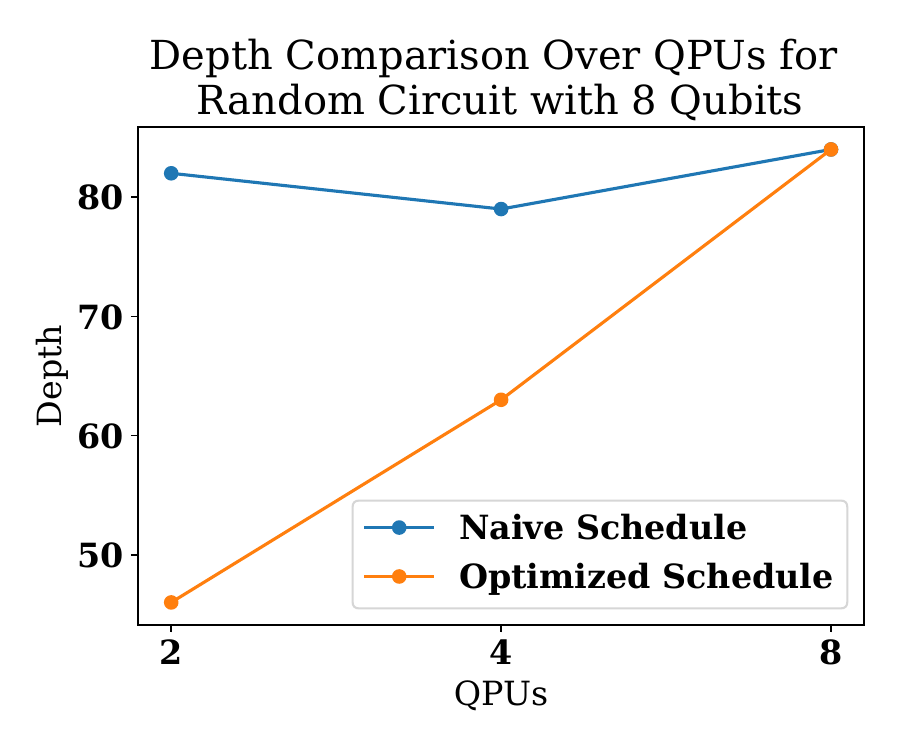}
        \caption{Random 8 qubit depth comparison.}
        \label{fig:random_8_depth}
    \end{subfigure}
    \hfill
    \begin{subfigure}[t]{0.3\textwidth}
        \centering
            \includegraphics[scale=0.35]{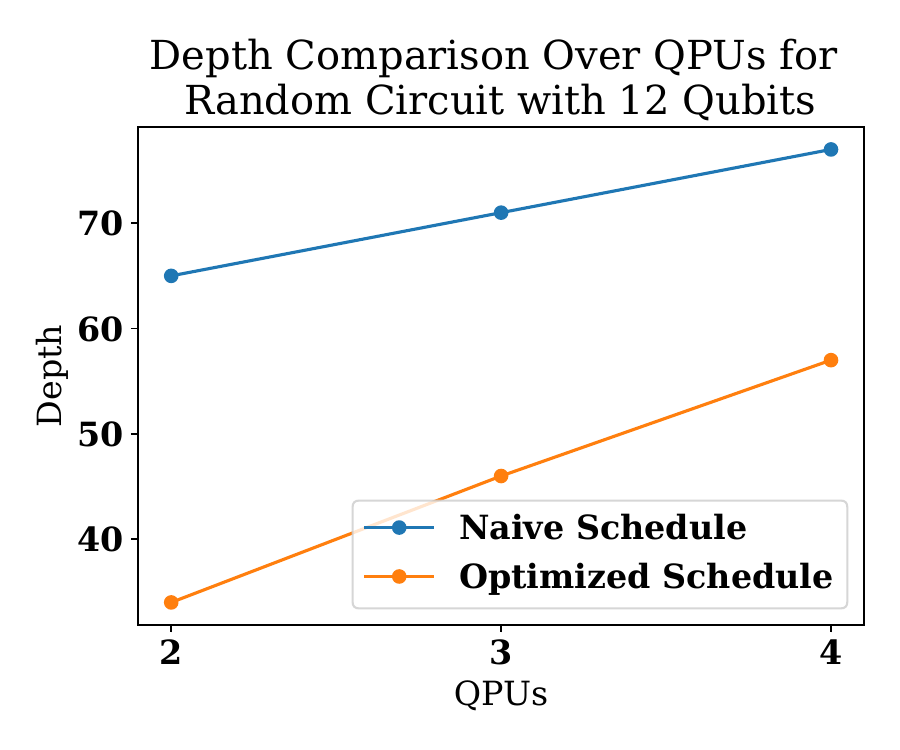}
        \caption{Random 12 qubit depth comparison.}
        \label{fig:random_12_depth}
    \end{subfigure}
    \caption{Comparison of the depth of the distributed circuits over different number of QPUs.}
    \label{fig:dqc_depth_comparison}
\end{figure*}

\section{Results}\label{sec:results}
The results of the experiments on the Grover circuit are shown in Fig. \ref{fig:grover_result}. In this experiment, the performance of the monolithic circuit was compared to a distributed version for two QPUs where each had two computational and two communication qubits. The monolithic circuit achieves a fidelity to the ideal state of 0.87 under noise while the distributed circuit is significantly lower at 0.70. The mean performance is identical for both schedules, i.e., optimizing the schedule using GP does not yield an improvement over a naive schedule.
The GHZ results are shown in Fig. \ref{fig:ghz_result}. The monolithic circuit achieves a fidelity of 0.97 and the distributed circuit for two QPUs a slightly lower value of 0.95. Although increasing the QPUs to four and eight further decreases the fidelity to 0.92 and 0.85 respectively with both schedules achieving an identical mean performance for two and four QPUs. With eight QPUs, the naive schedule performs slightly better. 
Fig. \ref{fig:vqc_8_results} presents the results for the VQC with eight qubits using network configurations with two, four, and eight QPUs. The monolithic circuit achieves a fidelity of 0.87 with the distributed circuits having a fidelity of 0.76, 0.66, and 0.51 for the respective network configuration. Both scheduling methods achieve similar results, only different in the eight QPU case where the optimized schedule is again slightly worse than the naive approach. 
Both the schedule and the number of QPUs influence the number of remote operations required; this in turn affects the depth of the circuit. Fig. \ref{fig:dqc_depth_comparison}
shows the depth for the VQC and random circuits used in the experiments. For the VQC circuit with eight qubits, shown in Fig. \ref{fig:vqc_depth}, the depth is the same for both naive and optimized schedules. For the random circuit with eight qubits (Fig. \ref{fig:random_8_depth}), the optimized schedule results in a significantly shallower circuit in the two and four QPU cases. In the eight QPU configuration the depth is identical to the naive schedule. In the 12 qubit random circuit, the optimized schedule results in a more efficient circuit in all cases (fig. \ref{fig:random_12_depth}). 
Results of the experiments on the eight qubit random circuit are shown in Fig. \ref{fig:random_8_results}. The monolithic circuit achieves a fidelity of 0.88 whereas the distributed versions with two, four, and eight QPUs result in a fidelity of 0.72, 0.7, and 0.68 respectively in the naive schedule experiments. With the optimized schedule, the distributed circuits achieve a fidelity of 0.84, 0.77, and 0.67. The difference between using a naive and optimized schedule are particularly evident for the two and four QPU cases (0.72 vs. 0.84 and 0.70 vs. 0.77).
In the next experiment, shown in Fig. \ref{fig:random_12_results}, a 12 qubit random circuit is evaluated using three different network configurations. The monolithic circuit has a fidelity of 0.86, while the distributed versions with the naive schedule achieve 0.75 (two QPUs), 0.71 (three QPUs), and 0.69 (four QPUs). With the optimized schedule, the distributed circuits achieve a fidelity of 0.82, 0.8, and 0.76, which is a vastly better performance than with the naive schedule.
Fig. \ref{fig:comm_qubit_results} shows the results of the experiment with a random circuit containing 12 qubits. For this experiment, the circuit was run for three different configurations, where only the number of communication qubits in each QPU is changed. The monolithic circuit achieves a fidelity of 0.86 and the distributed circuits with one, two, and three communication qubits and a naive schedule achieve 0.69, 0.68, and 0.68 respectively. With an optimized schedule, the fidelities were 0.76, 0.75, and 0.76, which again is significantly better than with the naive schedule. In these experiments, the influence of the number of communication qubits on the performance is not significant.
The results for the VQC with 12 qubits are depicted in Fig. \ref{fig:vqc_12_results}. The monolithic circuit achieves a fidelity of 0.81 while the distributed circuit with a naive schedule 0.7, 0.64, and 0.61. With the optimized schedule, the distributed circuit achieves fidelities of 0.71, 0.65, and 0.61.

\section{Discussion and Conclusion}\label{sec:discussion}
Remote operations consume entanglement, a valuable resource in quantum networks and necessary for DQC. Establishing and maintaining entanglement between distant nodes is costly, and therefore scheduling operations requiring quantum communication between QPUs should be optimized to avoid unnecessary consumption of expensive resources. Moreover, adding remote operations may increase the susceptibility to noise. 
In this work, the objective was to evaluate how distributed circuits with remote CX gates perform under noise in various configurations. We have shown that with an increasing number of QPUs the fidelity decreases. Additionally, we have seen that the qubit scheduling, i.e., the assignment of qubits to QPUs, also has an impact on the performance. However, this was not always the case, as experiments on VQC and GHZ circuits have shown; here both approaches achieved similar results. Both of these circuits have a similar entanglement pattern, i.e., CX gates are between neighboring qubits. In such circuit architectures, a naive qubit assignment as described above may be sufficient. Scheduling for different circuit architectures is a topic that should be explored in future work, that is, it could be worthwhile to investigate circuit patterns and their ideal assignment to QPUs. 
The noise model and the DQC simulation framework were deliberately simplistic in this study. The model is sufficient to investigate how the protocol affects the performance in different scenarios and configurations. However, in future work, more realistic network models that incorporate network topologies, entanglement generation, and repeaters should be explored. 

DQC is a promising paradigm for scaling quantum computing to allow the execution of large circuits that no QPU is able to run on their own. QPUs must still interact with each other through the use of communication protocols, which consume costly resources in the form of entanglement that must be generated, distributed, and maintained throughout the network. In this work, we evaluated how distributed circuits using the remote CX protocol perform under noise in various network configurations. Using a simplified noise and communication model, we showed that with increasing number of QPUs, the fidelity degrades, in some cases drastically to unacceptable fidelities. We additionally showed when and how qubit scheduling, that is, the assignment of qubits to QPUs affects the performance. The study provides initial results and insights, however, future work should consider more realistic network models and investigate pathways to allow the execution of distributed circuits without drastically degrading fidelity.

\bibliographystyle{IEEEtran}
\bibliography{IEEEabrv,bibliography}

\end{document}